\documentclass[final]{aipproc}
\layoutstyle{6x9}

\def\vec#1{\mathchoice{\mbox{\boldmath$\displaystyle#1$}}
{\mbox{\boldmath$\textstyle#1$}}
{\mbox{\boldmath$\scriptstyle#1$}}
{\mbox{\boldmath$\scriptscriptstyle#1$}}}
\makeatletter
\newcommand\erfc{\mathop{\operator@font erfc}\nolimits}
\def\slashchar#1{\setbox0=\hbox{$#1$}
   \dimen0=\wd0 \setbox1=\hbox{/} \dimen1=\wd1
   \ifdim\dimen0>\dimen1 \rlap{\hbox to \dimen0{\hfil/\hfil}} #1
   \else  \rlap{\hbox to \dimen1{\hfil$#1$\hfil}} / \fi}

\makeatother

\usepackage{graphicx}
\include{amssym}

\begin{document}

\title{NJL with eight quark interactions: \\ Chiral phases at finite
$T$
}



\author{B. Hiller}{address={Departamento de Física, Universidade
        de Coimbra, P-3004-516 Coimbra, Portugal}}

\author{A. A. Osipov}{address={Joint Institute for Nuclear Research,
        Laboratory of Nuclear Problems, 141980 Dubna, 
        Russia}}

\author{J. Moreira}{address={Departamento de Física, Universidade de
        Coimbra, P-3004-516 Coimbra, Portugal}}

\author{A. H. Blin}{address={Departamento de Física, Universidade de
        Coimbra, P-3004-516 Coimbra, Portugal}}

\begin{abstract}
The thermodynamic potential and thermal dependence of low lying mass
spectra of scalars and pseudoscalars are evaluated in a generalized
Nambu --- Jona-Lasinio model, which incorporates eight-quark
interactions. These are necessary to stabilize the scalar
ef\mbox{}fective potential for the light and strange quark
f\mbox{}lavors, which would be otherwise unbounded from below. In
addition it turns out that they are also crucial to i) lower the
temperature of the chiral transition, in conformity with lattice
calculations, ii) sharpen the temperature interval in which the
crossover occurs, iii) or even allow for first order transitions to
occur with realistic quark mass values, from certain critical values
of the parameters. These are unprecedented results which cannot be
obtained within the NJL approaches restricted to quartic and six-quark
interactions.
\end{abstract}

\maketitle

Nambu -- Jona-Lasinio (NJL) models (\cite{Nambu:1961}, for reviews see
e.g. \cite{Klevansky:1992,Hatsuda:1994}) have the very appealing
property of describing dynamical breakdown of chiral symmetry. In the
present talk we show our recently obtained results \cite{Osipov:2008a},
drawing particular attention to two distinct patterns of chiral
symmetry breaking (SB) and their impact on the nature, temperature
values and slopes of chiral transitions. This study has its roots in
the observation \cite{Osipov:2006a} that extensions of the
NJL model to accomodate the approximate $SU(3)$ flavor symmetry of the
$u,d,s$-quarks and the $U_A(1)$ breaking instanton induced 't Hooft
interaction \cite{Hooft:1976}, display an unstable/ metastable vacuum
in stationary phase (SPA)/ mean field approximations, and subsequent
resolution of this problem by the addition of eight-quark interactions
to the Lagrangian \cite{Osipov:2006,Osipov:2007a}. The multi-quark interactions considered are the
most general non-derivative chiral symmetric spin zero combinations.
A set of stabilization conditions constrain the coupling strengths,
from which the $N_c$ dependence of the OZI-violating eight-quark
interactions is inferred. Furthermore SPA coincides then with the mean
field approach. The characteristics of the low lying pseudoscalar and
scalar nonets at $T=0$ have been reevaluated in the present framework
\cite{Osipov:2007a}. One main conclusion is that identical spectra,
except for the scalar singlet-octet mixing channel (strongest
effect on the $\sigma$-meson mass) can be obtained from two
distinct ef\mbox{}fective potentials. They are generated by just
changing the strengths of 4- and 8-quark (q) couplings keeping all
other model parameters fixed.
The 4q
coupling regulates the curvature at the origin and thus determines either
the Wigner-Weyl or broken phase (in absence of other interactions).
Higher order interactions can however induce symmetry breaking on top
of the Wigner-Weyl phase, i.e. a second minimum arises with a finite
condensate while the origin keeps further its status as a minimum.
This is in contrast with the case in which the curvature at the origin
represents a maximum of the effective potential. Then higher order
interactions will not generate further minima, but simply shift the
position of the existing minimum. Here their action results in a
perturbative effect around the broken phase, while in the former case
they are the motor for non-perturbative SB. These starting
configuartions with double vacua at $T=0$ lead to a lowering of the
critical temperature. The phenomenom of multiple vacua is
known to occur within several approaches to the QCD vacuum
\cite{Bicudo:2006}. We show below that these patterns of SB are still
present for realistic values of quark masses, although the origin
loses of course its significance as a reference point for the
curvature. Explicit fits reveal further that it is the 't Hooft 6q
strength and not the 4q or 8q couplings which induce SB. This does of
course not preclude the important role played by these interactions,
without which neither stability of the vacuum nor a "twin fit" of mass
spectra is possible.

The present analysis focuses on the gap equations at finite temperature,
whose solutions represent the extrema of the thermodynamic potential.
The expressions were derived within a generalized heat kernel scheme
\cite{Osipov:2001} which takes into account quark mass dif\mbox{}ferences
in a symmetry preserving way at each order of the expansion at $T=0$
in \cite{Osipov:2006} and we will consider from now on the isospin
limit, $m_u=m_d\ne m_s$
\vspace{-0.2cm}

\begin{equation}
\label{gap}
   h_u+\displaystyle\frac{N_c}{6\pi^2}\, M_u
       \left(3I_0-\Delta_{us} I_1 \right)=0, \qquad
   h_s+\displaystyle\frac{N_c}{6\pi^2}\, M_s
       \left(3I_0+2\Delta_{us} I_1 \right)=0,
\end{equation}
\vspace{-0.3cm}

\noindent which must be solved selfconsistently with the stationary phase
equations
\vspace{-0.2cm}

\begin{equation}
\label{SPA}
    Gh_i + \Delta_i +\frac{\kappa}{16}\, h_jh_k
         + \frac{g_1}{4}\, h_i (h_i^2+h_j^2+h_k^2)
         + \frac{g_2}{2}\, h_i^3=0.
\end{equation}
\vspace{-0.3cm}

Here $\Delta_{ij}=M_i^2-M_j^2$, $\Delta_i=M_i-m_i$, $i,j,k=u,d,s$
(with cyclic permutations of $u,d,s$ for three possible equations) and
$i\neq j\neq k$, $M_i$ denote the constituent quark masses. It is
obvious that $h_u=h_d$ for the considered case. The factors $I_i$ are
given by the average
\vspace{-0.2cm}

\begin{equation}
   I_i =\frac{1}{3}\left[2J_i(M_u^2)+J_i(M_s^2)\right], \quad
   J_i(M^2)=\int\limits_0^\infty\frac{{\rm d}t}{t^{2-i}}\,\rho
   (t\Lambda^2)\exp[-t M^2],
\end{equation}
\vspace{-0.3cm}

\noindent and represent one-quark-loop integrals with the Pauli-Villars
regularization kernel \cite{Pauli:1949} $\rho(t\Lambda^2)=1-(1+t
\Lambda^2)\exp[-t \Lambda^2]$, where $\Lambda$ is an ultraviolet
cutoff (the model is not renormalizable). For this case one needs only
to know
\vspace{-0.5cm}

\begin{equation}
\label{j0}
   J_0(M^2)=\Lambda^2- M^2\ln\left(1+\frac{\Lambda^2}{M^2}\right),
   \quad
   J_1(M^2)=\ln\left(1+\frac{\Lambda^2}{M^2}\right)
      -\frac{\Lambda^2}{\Lambda^2+M^2}\ .
\end{equation}
\vspace{-0.3cm}

The model parameters are the four quark coupling $G\sim N_c^{-1}$, the
't Hooft interaction coupling $\kappa \sim N_c^{-3}$, the  eight-quark
couplings $g_1,g_2$ ($g_1$ multiplies the OZI violating combination),
the current quark masses $m_i$ and the cutoff $\Lambda$. The stability
of the effective potential is guaranteed if the couplings fulfill the
following inequality \cite{Osipov:2006}
\vspace{-0.2cm}

\begin{equation}
\label{ineq1}
   g_1>0, \quad g_1 +3g_2>0, \quad
   G>\frac{1}{g_1}\left(\frac{\kappa}{16}\right)^2,
\end{equation}
\vspace{-0.3cm}

from which we deduce that $g_1$ must scale at most as $N_c^{-5}$ and
at least as $N_c^{-4}$ \cite{Osipov:2007a}.

The generalization to finite temperature of these expressions occurs in
the quark loop integrals $J_0,J_1$. After introducing the Matsubara
frequencies \cite{Matsubara}

\vspace{0.3cm}

\noindent{\small TABLE 1:
Parameters of the model at $T=0$. The couplings have the following
units: $G$ (GeV$^{-2}$), $\kappa$ (GeV$^{-5}$), $g_1,\, g_2$
(GeV$^{-8}$), $m_u=m_d, m_s$, and $\Lambda$ are given in MeV.
The values of constituent quark masses $M_u=M_d$ and $M_s$
are shown in MeV (only the case of global minima).} \\[0.1cm]
\noindent
\begin{tabular}{lrrrrrrrrr}
\hline
Sets &\ \ \ \ \ \ $m_u$  &\ \ \ \ \ $\ m_s$  &\ \ \ \ $\ M_u$
     &\ \ \ \ $\ M_s$ &\ \ \ \ \ $\ \Lambda$  &\ \ \ \ \ \ $\ G$
     &\ \ \ \ \ $-\kappa$ &\ \ \ \ \ \ \ \ \ \ $g_1$
     &\ \ \ \ \ $\ g_2$  
\\
\hline
a  & 5.8  & 183 & 348 & 544 & 864  & 10.8  & 921   & 0*      &0*   \\ 
b  & 5.8  & 181 & 345 & 539 & 867  & 9.19  & 902   & 3000*   &-902 \\ 
c  & 5.9  & 186 & 359 & 544 & 851  & 7.03  & 1001  & 8000*   &-47  \\ 
d  & 5.8  & 181 & 345 & 539 & 867  & 5.00  & 902   & 10000*  &-902 \\ 
\hline
\end{tabular}
\vspace{0.2cm}

\vspace{0.3cm}

\noindent{\small TABLE 2: The masses, weak decay constants of light
pseudoscalars (in MeV), the singlet-octet mixing angle $\theta_p$ (in
degrees), and the quark condensates $\langle\bar uu\rangle,
\langle\bar ss\rangle$ expressed as usual by positive combinations in
MeV.}  \\[0.1cm]
\noindent
\begin{tabular}{lrrrrrrrrr}
\hline
Sets &\ \ \ \ \ \ $m_\pi$ &\ \ \ \ \ \ $m_K$ &\ \ \ \ \ $m_\eta$
     &\ \ \ \ \ $m_{\eta'}$ &\ \ \ \ $f_\pi$ &\ \ \ \ $f_K$
     &\ $\theta_p $  &$-\langle\bar uu\rangle^{\frac{1}{3}}$
     &$-\langle\bar ss\rangle^{\frac{1}{3}}$ \\
\hline
a  &138*  &494*  &480  &958*  &92*  &118*  & -13.6  &237  &191  \\
b  &138*  &494*  &480  &958*  &92*  &118*  &-13.6   &237  &192  \\
c  &138*  &494*  &477  &958*  &92*  &117*  & -14.0  &235  &187  \\
d  &138*  &494*  &480  &958*  &92*  &118*  & -13.6  &237  &192  \\
\hline
\end{tabular}
\vspace{0.2cm}

\vspace{-0.2cm}

\begin{equation}
\label{j0t}
    J_0(M^2)\rightarrow J_0(M^2,T)=16\pi^2 T\!
    \sum_{n=-\infty}^{\infty}
    \int\!\frac{{\rm d}^3{p}}{(2\pi )^3}
    \int\limits_0^\infty\!{\rm d}s\, \rho (s\Lambda^2)
    e^{-s[(2n+1)^2\pi^2 T^2+ \vec{p}^2+M^2]},
\end{equation}
\vspace{-0.5cm}

\noindent and using the Poisson formula
\vspace{-0.2cm}

\begin{equation}
   \sum_{n=-\infty}^{\infty} F(n) = \sum_{m=-\infty}^{\infty}
   \int_{-\infty}^{+\infty} {\rm d}x\, F(x) e^{i2\pi mx},
\end{equation}
\vspace{-0.3cm}

\noindent where $F(n)=\exp [-s(2n+1)^2\pi^2 T^2]$, one integrates over the
3-momentum $\vec{p}\,$ leading to
\vspace{-0.6cm}

\begin{equation}
\label{j0t2}
   J_0(M^2,T)=\int\limits_0^\infty
   \frac{{\rm d}s}{s^2}\,\rho (s\Lambda^2) e^{-s M^2}
   \left[1+2\sum_{n=1}^{\infty} (-1)^{n}
   \exp\left(\frac{-n^2}{4sT^2}\right)\right].
\end{equation}
\vspace{-0.3cm}

Similarly one gets $J_1 (M^2,T) = -\partial J_0 (M^2,T)/\partial M^2.$
One recovers at $T=0$ the starting expressions (\ref{j0}) and
verifies also that $\lim_{T\to\infty}J_{0,1}(M^2,T)=0$.

Using these formulae in (\ref{gap}), we solve the system
(\ref{gap})-(\ref{SPA}) numerically, assuming that the model
parameters $G, \kappa , g_1, g_2, m_i, \Lambda$ do not depend on
the temperature. As a result we obtain the temperature dependent
solutions $M_i(T)$, representing the extrema of the thermodynamic
potential.

\begin{figure}[t]
\includegraphics[height=4.4cm,angle=0]{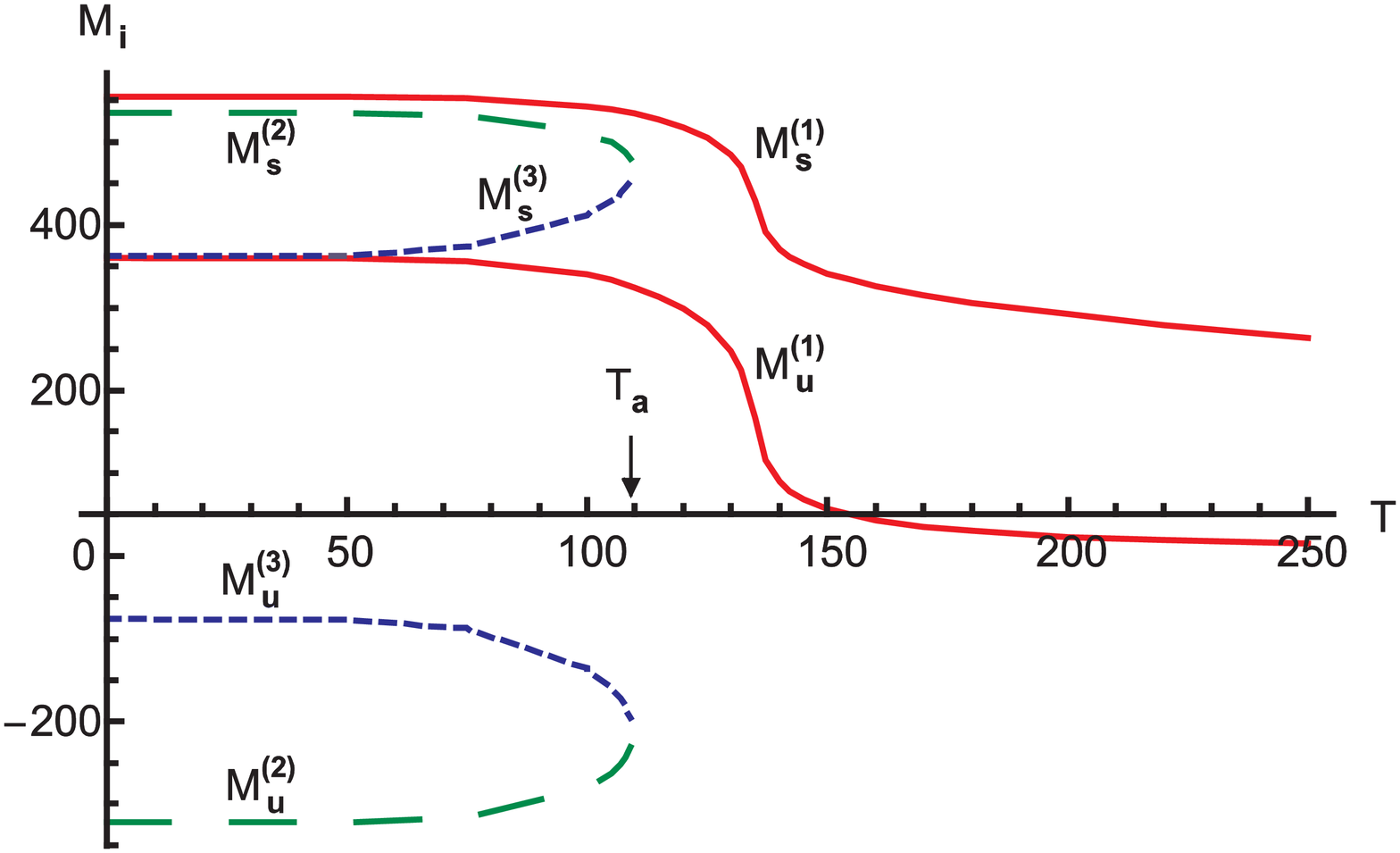}\hspace{0.1cm}
\includegraphics[height=4.4cm,angle=0]{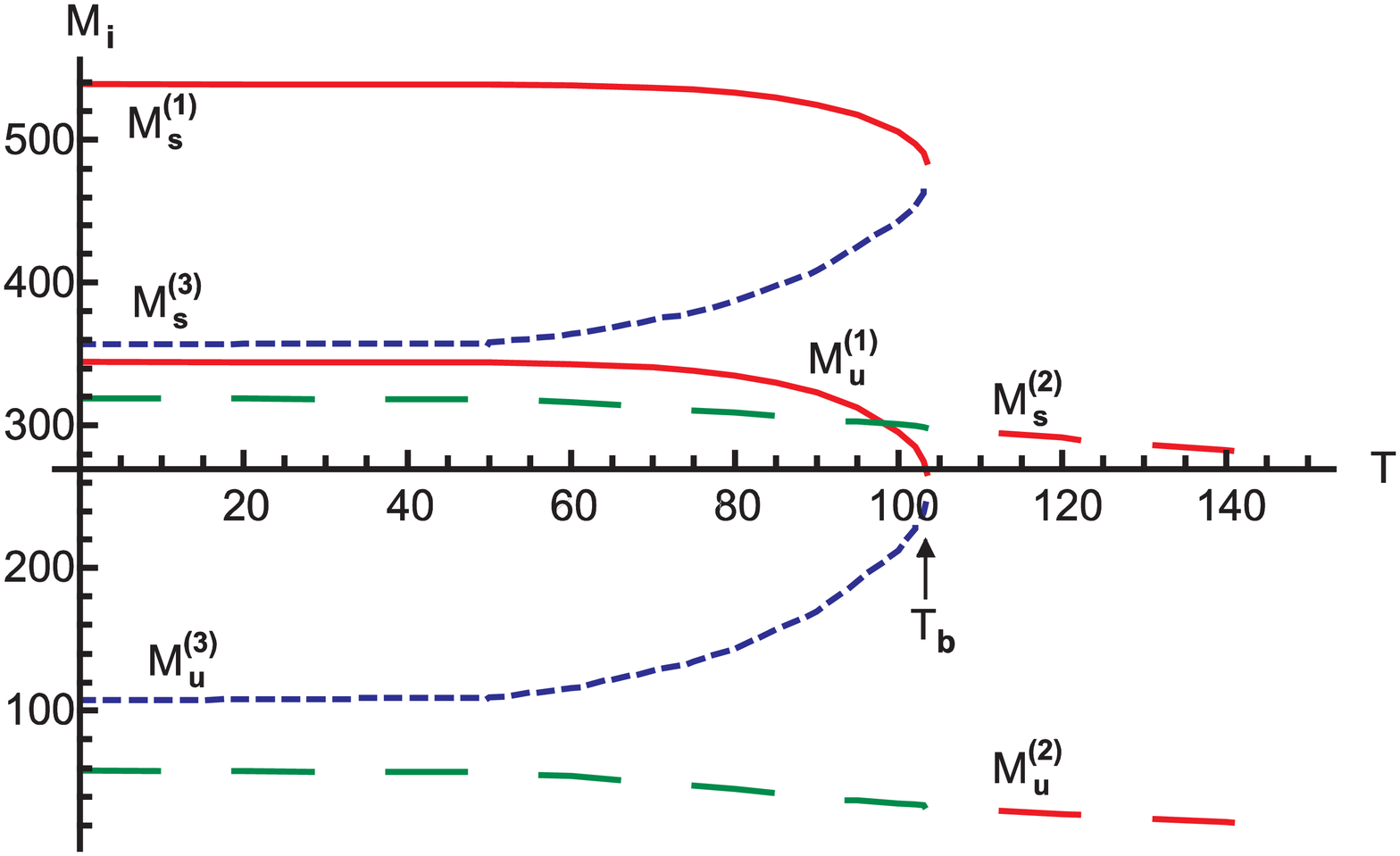}
\caption{{\it Left}: Branches of $M_u^{(i)}(T),M_s^{(i)}(T)$ pairs, denoting extremal
         points of the thermodynamic potential, given by the solutions
         of the gap equations as functions of the temperature. Solid
         lines (start at $T=0$ as deepest minima), dashed lines (start
         as relative minima at $T=0$) and dotted lines (saddle at
         $T=0$) for the parameter set (c). Only one branch is in the
         physical (positive mass) region (solid curves).
         {\it Right}: The same as on the left but for set (d). All
         branches lie in the physical region and coexist up to
         $T=T_b$, from this value on only one branch survives, with
         much lower values of $M_i$.}
\label{fig1}
\end{figure}

The fit of the model parameters is obtained by fixing low lying
pseudoscalar and scalar meson characterisitics at $T=0$, (stars denote
input) in Tables 1-3. As already observed in \cite{Osipov:2007a},
$f_0(600)$ is the main observable responsive to changes in the OZI
violating eight-quark interaction term, diminishing with increasing
strength of the $g_1$ coupling. The $M_i(T)$ are shown in Fig.
\ref{fig1} (sets c and d). There are either one or three $(M_u^{(i)}
= M_d^{(i)}, M_s^{(i)})$, $i=1,2,3$ couples of solutions at
f\mbox{}ixed values of $T$. For set (c) (as well as (b)) only one
branch of solutions is physical, {\it i.e.} positive valued. The other
two have negative values for the light quark masses. One sees however
(set c) that the onset of the transition occurs at a value of $T=T_a$
for which the other unphysical two branches meet and cease to exist.
The rapid crossover occurs in the short temperature interval
$125<T<140$ MeV. The crossover pattern is in contrast with the $SU(3)$
limit case with zero current quark masses, where one branch collapses
to the origin $M_u=M_d=M_s=0$ for all values of the remaining model
parameters and $T$. In this case the transition is first order
\cite{Osipov:2007b}.

We observe however that below a certain critical value of
$G\Lambda^{-2}$ (accompanied by a critical value of $g_1$) one
obtains, also for the case of realistic quark masses, solutions with
all branches positive valued at any $T$. This is the case shown in Fig.
\ref{fig1} set (d). Two of the branches (starting from the stable
minimum and the saddle solution at T=0) merge in the physical region
at a certain $T_b$ and the surviving branch has a significantly lower
mass value. This leads to the discontinuities in observables typical
of first order transitions. The decrease in temperature observed in
sets (c) and (d) is welcome in view of recent lattice calculations
\cite{Aoki:2006}, obtained for finite values of the quark masses. In
this case there is evidence that a rapid crossover occurs, as opposed
to an expected first order transition for the massless case
\cite{Wilczek:1984}-\cite{Lenaghan:2000}. Lattice QCD data have not
unambiguously settled the question about the order of the chiral
transition. For physical values of the quark masses, calculations with
staggered fermions favor a smooth crossover transition
\cite{Brown:1990}, while calculations with Wilson fermions predict the
transition to be first order \cite{Iwasaki:1996}. At zero
chemical potential there is growing evidence that the transition is
crossover, which would set an upper bound for the OZI-violating 8q
coupling $g_1$.

We finally remark that the parameter set (a), without eight-quark
interactions, evolves as function of $T$ qualitatively as in Fig. 1,
however the crossover takes place at much larger temperatures,
$T\simeq 210$ MeV. Also the transition is much smoother than for set
(c).

\begin{figure}[t]
\hspace{0.5cm}\includegraphics[height=4.5cm,width=10cm,angle=0]{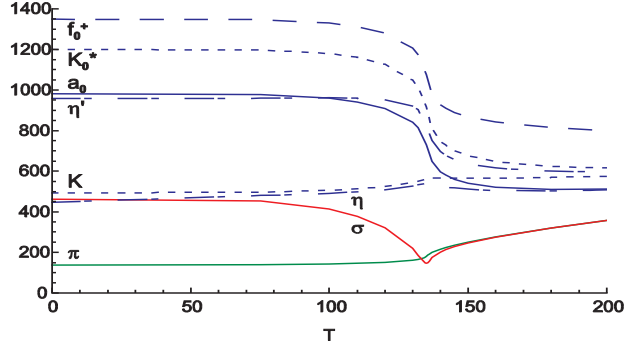}
\caption{The masses of pion, $\sigma=f_0(600),\eta$, kaon,$\eta'$, $a_0$, $K^*_0$ and
         $f_0(980)$ from bottom to top, for set (c), as functions of T (all in MeV).
         }
\label{fig3}
\end{figure}

The masses of scalar and pseudoscalar mesons at f\mbox{}inite
temperature obtained for the set (c) are shown in Fig. \ref{fig3}. As
can be seen there is a rapid crossover for all meson masses in the
same temperature interval as in Fig. \ref{fig1}. However, neither this
rapid crossover nor the f\mbox{}irst order transition case (d) do
imply restoration of chiral or $U_A(1)$ symmetry, but only the
recovery of a distorted Wigner -- Weyl phase, with the minimum of the
thermodynamic potential shifted to f\mbox{}inite quark mass values due
to f\mbox{}lavor breaking ef\mbox{}fects.

The role played by the dif\mbox{}ferent multi-quark interactions can
be further understood by analyzing the following two limits, with the
parameter set (c) as starting condition.

Case 1: We set $g_1=g_2=\kappa=0$ and remaining parameters as in (c).
In this limit the gap equation has only one solution for the
considered parameter set, thus the system is in a distorted
Wigner -- Weyl phase,

Case 2: We set $\kappa=0$ and all other parameters fixed as in (c).
In this case there is no $U_A(1)$ breaking, but OZI violating
ef\mbox{}fects are present. We verify that in this limit the gap
equation has also only one solution, being again in a distorted
Wigner -- Weyl phase.

Thus the spontaneous symmetry breakdown seen in the full set (c) at
$T=0$ (and also in sets b and d) is driven exclusively by the 't Hooft
interaction strength $\kappa$. We wish not to include case (a) in the
present discussion, as it violates the stability conditions of
(\ref{ineq1}).

\vspace{0.3cm}

\noindent{\small TABLE 3: The masses of the scalar nonet (in MeV) at
$T=0$, and the corresponding singlet-octet mixing angle $\theta_s$ (in
degrees).} \\[0.1cm]
\noindent
\begin{tabular}{lrrrrr}
\hline
   Sets &\ \ \ \ \ \ \ \ \ $\ m_{a_0 (980)}$
        &\ \ \ \ \ \ \ \ \ $\ m_{K_0^* (800)}$
        &\ \ \ \ \ \ \ \ \ $\ m_{f_0(600)}$
        &\ \ \ \ \ \ \ \ \ $\ m_{f_0(980)}$
        &\ \ \ \ \ \ \ \ \ \ $\ \ \ \theta_s$  \\
\hline
a    &963.5   &1181   &707   &1353   &24  \\
b    &1024*   &1232   &605   &1378   &20  \\
c    &980*    &1201   &463   &1350   &24  \\
d    &1024*   &1232   &353   &1363   &16  \\
\hline
\end{tabular}
\vspace{0.3cm}

In conclusion, the present study indicates that chiral eight-quark
interactions have a strong ef\mbox{}fect on the temperature dependence
of observables described by NJL models, offering a plethora of solutions
deeply rooted in the nature of chiral transitions. Within the model
its origins can be traced back to the pattern of dynamical chiral symmetry
breaking. We have discussed at length how the different patterns emerge.
Mesonic spectra built on the spontaneously broken vacuum
induced by the 't Hooft interaction strength, as opposed to the
commonly considered case driven by the four-quark coupling, undergo a
rapid crossover to the unbroken phase, with a slope and at a
temperature which is regulated by the strength of the OZI violating
eight-quark interactions. This strength can be adjusted in consonance
with the four-quark coupling and leaves the spectra unchanged, except
for the sigma meson mass, which decreases. This ef\mbox{}fect also
explains why in the crossover region the sigma meson mass drops
slightly below the pion mass. A first order transition behavior
is also a possible solution within the present approach. Additional
information from lattice calculations and phenomenology is necessary
to f\mbox{}ix f\mbox{}inally the strength of interactions. We expect
that the role of eight-quark interactions are of equal importance in
studies involving a dense medium and extensions of the model with the
Polyakov loop \cite{Polyakov:1978}.    
The latter is known to increase the transition temperature by $\sim
25$ MeV \cite{Aoki:2006}. In relation with the NJL model the role of the
Polyakov loop has been investigated in several papers, see e.g
\cite{Fukushima:2004}. 
The present study can be
extended likewise.  In the two flavor NJL the inclusion of eight quark
interactions has been analyzed in connection with finite $T$, chemical
potential and Polyakov loop \cite{Kashiwa:2006}, where the relevance
of eight-quark interactions has been reported. Eight-quark physics has
further been explored in presence of a constant magnetic field
\cite{Osipov:2008c}, and also in that case it provides for a rich structure
of the effective potential.
\vspace{0.2cm}

This work has been partly supported by grants of Fundação para a
Ci\^encia e Tecnologia, POCI 2010 and FEDER, POCI/FP/63930/2005,
POCI/FP/81926/2007.


\vspace{-0.2cm}

\begin{thebibliography}{47}
\expandafter\ifx\csname natexlab\endcsname\relax\def\natexlab#1{#1}\fi
\providecommand{\enquote}[1]{``#1''}
\expandafter\ifx\csname url\endcsname\relax
  \def\url#1{\texttt{#1}}\fi
\expandafter\ifx\csname urlprefix\endcsname\relax\def\urlprefix{URL }\fi
\providecommand{\eprint}[2][]{\url{#2}}

\bibitem{Nambu:1961} Y. Nambu and G. Jona-Lasinio, Phys. Rev. {\bf 122},
    345 (1961); {\bf 124}, 246 (1961); V. G. Vaks and A. I. Larkin,
    Zh. \'{E}ksp. Teor. Fiz. {\bf 40}, 282 (1961) [Sov. Phys. JETP
    {\bf 13}, 192 (1961)].
\bibitem{Klevansky:1992} S.P. Klevansky, Rev. Mod. Phys. {\bf 64},
    649 (1992).
\bibitem{Hatsuda:1994} T. Hatsuda, T. Kunihiro, Phys. Rep. {\bf 247},
    221 (1994).
\bibitem{Osipov:2008a} A.A. Osipov, B. Hiller, J. Moreira, A.H. Blin,
    Phys. Lett. B {\bf 659}, 270 (2008)
\bibitem{Osipov:2006a} A.A. Osipov, B. Hiller, V. Bernard, A.H. Blin,
    Ann. of Phys. {\bf 321}, 2504 (2006).
\bibitem{Hooft:1976} G. 't Hooft, Phys. Rev. D {\bf 14}, 3432 (1976);
    G. 't Hooft, Phys. Rev. D {\bf 18}, 2199 (1978).
\bibitem{Osipov:2006} A.A. Osipov, B. Hiller, J. da
    Provid\^encia, Phys. Lett. B {\bf 634}, 48 (2006).   
\bibitem{Osipov:2007a} A.A. Osipov, B. Hiller, A.H. Blin, J. da
    Provid\^encia, Ann. of Phys. {\bf 322}, 2021 (2007). 
\bibitem{Bicudo:2006} P. Bicudo, J. E. Ribeiro, A. V. Nefediev,
    Phys. Rev. D {\bf 65}, 085026 (2002).
\bibitem{Osipov:2001} A. A. Osipov, B. Hiller, Phys. Lett. B {\bf
    515}, 458 (2001); Phys. Rev. D {\bf 64}, 087701 (2001);
    Phys. Rev. D {\bf 63}, 094009 (2001).
\bibitem{Pauli:1949} W. Pauli and F. Villars, Rev. Mod. Phys. {\bf
    21}, 434 (1949).
\bibitem{Matsubara} J. I. Kapusta, ``Finite-Temperature Field
    Theory'', Cambridge: Cambridge University Press, 1989.
\bibitem{Osipov:2007b} A. A. Osipov, B. Hiller, J. Moreira,
    A. H. Blin, J. da Providencia, Phys. Lett. B {\bf 646}, 91 (2007).
\bibitem{Aoki:2006} Y. Aoki, G. Endrodi, Z. Fodor, S. D. Katz,
    K. K. Szabo, Nature {\bf 443}, 675 (2006);
    Y. Aoki, G. Endrodi, Z. Fodor, S. D. Katz,
    K. K. Szabo,  Phys. Lett. B {\bf 643}, 46 (2006).
\bibitem{Wilczek:1984} R. D. Pisarski, F. Wilczek, Phys. Rev. D
    {\bf 29}, 338 (1984).
\bibitem{Brown:1990} F.R. Brown {\it et al.,} Phys. Rev. Lett. {\bf
    65}, 2491 (1990).
\bibitem{Ortmanns:1996} H. Meyer-Ortmanns, Rev. Mod. Phys. {\bf 68},
    473 (1996).
\bibitem{Lenaghan:2000} J.T. Lenaghan, D.H. Rischke,
    J. Schaffner-Bielich, Phys. Rev. D {\bf 62}, 085008 (2000).
\bibitem{Iwasaki:1996} Y. Iwasaki, K. Kanaya, S. Kaya, S. Sakai,
        and T. Yoshie, Zeit. Phys. C {\bf 71}, 343 (1996).
\bibitem{Polyakov:1978} A. Polyakov,
    ICTP report IC/78/4 (1978), unpublished;
    G. 't Hooft, Nucl. Phys. B {\bf 153}, 14 (1979);
    B. Svietitsky, Phys. Rep. {\bf 132}, 1 (1986).
\bibitem{Fukushima:2004} K. Fukushima, Phys. Lett. B {\bf 591},
    277 (2004);
    E. Megias, E. Ruiz Arriola, L. L. Salcedo,
    Phys. Rev. D {\bf 74}, 065005 (2006); 
    AIP
    {\bf 892}, 444 (2007);
    H. Hansen, et al.,
    Phys. Rev. D {\bf 75}
    065004 (2007). 
\bibitem{Kashiwa:2006} K. Kashiwa, H. Kouno, T. Sakaguchi,
    M. Matsuzaki, M. Yahiro, Phys. Lett. B {\bf 647}, 446 (2007);
    K. Kashiwa, M. Matsuzaki, H. Kouno, M. Yahiro, Phys. Lett.
    B {\bf 657}, 143 (2007); Phys. Lett. B {\bf 662}, 26 (2008);
    arXiv:0803.1902 [hep-ph]
\bibitem{Osipov:2008c} A. A. Osipov, B. Hiller, A. H. Blin, J. da
    Provid\^encia, Phys. Lett B {\bf 650}, 262 (2007); SIGMA 4, 024
    (2008).
\end{thebibliography}
\end{document}